\documentclass[prb,twocolumn,showpacs,superscriptaddress,floatfix]{revtex4-1}
\usepackage{graphicx,amsfonts,amssymb,amsmath,hyperref}

\newif\ifhyper
\hypertrue
\ifhyper
\hypersetup{
  citecolor = {green},
  colorlinks = {true}, 
  urlcolor = {blue} 
}
\fi

\newlength{\ldag}
\begin{document}

\title{Self-duality and bound states of the toric code model in a transverse field}

\author{Julien Vidal}
\email{vidal@lptmc.jussieu.fr}
\affiliation{Laboratoire de Physique Th\'eorique de la Mati\`ere Condens\'ee,
CNRS UMR 7600, Universit\'e Pierre et Marie Curie, 4 Place Jussieu, 75252
Paris Cedex 05, France}

\author{Ronny Thomale}
\email{thomale@tkm.uni-karlsruhe.de}
\affiliation{Institut f\"{u}r Theorie der Kondensierten Materie, Universit\"{a}t Karlsruhe, D-76128 Karlsruhe, Germany}

\author{Kai Phillip Schmidt}
\email{schmidt@fkt.physik.uni-dortmund.de}
\affiliation{Lehrstuhl f\"ur Theoretische Physik I, Otto-Hahn-Stra\ss e 4, D-44221 Dortmund, Germany}

\author{S\'ebastien Dusuel}
\email{sdusuel@gmail.com}
\affiliation{Lyc\'ee Saint-Louis, 44 Boulevard Saint-Michel, 75006 Paris, France}


\begin{abstract}
We investigate the effect of a transverse magnetic field on the toric code model. We show that this problem can be mapped onto the Xu-Moore model and thus onto the quantum compass model which are known to be self-dual. We analyze  the low-energy spectrum by means of perturbative continuous unitary transformations and determine accurately the energy gaps of various symmetry sectors.   
Our results are in very good agreement with exact diagonalization data for all values of the  parameters except at the self-dual point where level crossings are responsible for a first order phase transition between a topological phase and a polarized phase. Interestingly, bound states of two and four quasiparticles with fermionic and bosonic statistics emerge, and display dispersion relations of reduced dimensionality.
 
 \end{abstract}

\pacs{71.10.Pm, 75.10.Jm, 03.65.Vf, 05.30.Pr}

\maketitle

%
%
%
%
Topologically ordered phases, such as those present in fractional quantum Hall systems~\cite{Tsui82,Laughlin83,Wen95}, have attracted much attention in the last few years. Indeed, in his seminal paper, Kitaev showed that topologically degenerate ground states may serve as a robust quantum memory~\cite{Kitaev03}, while braiding of anyonic 
excitations~\cite{Leinaas77,Wilczek82_1} can be used for fault-tolerant quantum computation.
Topologically protected qubits have left the realm of theory since superconducting nanocircuits led to their first experimental realization~\cite{Gladchenko09}. Recent progress in the field of ultracold atoms trapped in optical lattices also promises an implementation of such systems~\cite{Duan03,Micheli06,Jiang08}.

Although a topological quantum memory is, by nature, protected from decoherence, it is natural to wonder how large a local perturbation can be before this protection fails. 
With respect to this problematics, the toric code model (TCM),  which is undoubtedly one of the simplest model displaying topological order~\cite{Kitaev03}, is a perfect test ground. In the presence of parallel magnetic fields, the breakdown of the topological phase has been shown to be caused by single-anyon condensation \cite{Trebst07,Hamma08,Tupitsyn10,Vidal09_1}, leading to two second-order transition lines merging in a topological multicritical point~\cite{Tupitsyn10,Vidal09_1}.

The aim of this Rapid Communication is to investigate the influence of  a transverse field in the TCM which turns out to display completely different physics. Indeed, as we shall see, this model can be mapped onto the self-dual Xu-Moore model proposed to describe superconducting arrays~\cite{Xu04,Xu05}. Note that the Xu-Moore model can also be mapped onto the quantum compass model~\cite{Nussinov05_1} relevant for orbitally frustrated systems and for topologically protected qubits~\cite{Kugel82,Doucot05,Dorier05}.  
All results given below are thus also valid for these two models as far as the energy spectrum is concerned.
In the following, we compute the low-energy spectrum by means of perturbative continuous unitary transformations (PCUTs) and compare our results with exact diagonalization (ED) data. Our results reveal the existence of a first order phase transition at the self-dual point and emphasize the importance of  strong binding effects leading to a plethora of multi-quasiparticle bound states  with kinetics  of reduced dimensionality.

\begin{figure}[t]
 \includegraphics[width=0.4\columnwidth]{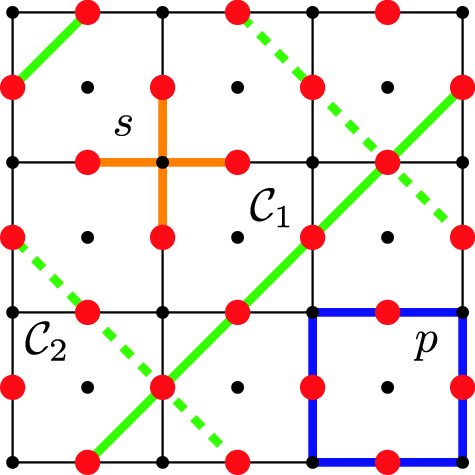}
 \caption{(Color online). Original square lattice on which plaquettes $p$ and
   stars $s$ are defined. Big red (small black) dots define the lattice
   $\Lambda$ ($\widetilde \Lambda$). Here, we show the lattice with $N=18$ spins and (implicit) periodic boundary conditions. Contour $\mathcal{C}_1$ ($\mathcal{C}_2$) is one of the diagonal (anti-diagonal) cycles used to define conserved
   parity operators.}
 \label{fig:1}
\vspace{-10pt}
\end{figure}

%
%
\emph{Model ---}
%
%
The transverse-field TCM Hamiltonian reads as
%
%
\begin{equation}
\label{eq:ham}
  H = -J \sum_{s} A_s - J \sum_{p} B_p - h_y \sum_i \sigma_i^y,
\end{equation} 
%
%
where $A_s=\prod_{i \in s}\sigma_i^x$, $B_p=\prod_{i \in p} \sigma_i^z$, and
the $\sigma_i^\alpha$'s are Pauli matrices. Subscripts $s$ and $p$ refer to
stars (vertices) and plaquettes of a square lattice, whereas $i$ runs over
all bonds where spin degrees of freedom are located (see Fig.~\ref{fig:1}).
%
%
\vspace{0pt}
%
%
In zero field, one recovers the TCM~\cite{Kitaev03} whose topological
ground state has eigenvalue +1 for all $A_s$ and $B_p$ operators. 
Excitations are $\mathbb{Z}_2$-charges with  eigenvalue $-1$ for one $A_s$
($\mathbb{Z}_2$-fluxes with  eigenvalue $-1$ for one $B_p$) localized
on the stars (plaquettes). These particles are hard-core bosons with mutual half-fermionic (semionic) statistics. Charges (or fluxes) can only
appear in pairs for a system with periodic boundary conditions.  In a magnetic
field, elementary excitations become dressed anyonic quasi-particles (QP)~\cite{Vidal09_1}.
In the opposite limit $J=0$, the ground state is fully polarized. Elementary
excitations are spin flips (magnons), which are likewise dressed when switching on $J$.

Although $A_s$ and $B_p$ are no longer conserved in a transverse field,
the parity operator $\prod_{i\in\mathcal{C}}\sigma_i^y$ still commutes with
$H$, provided $\mathcal{C}$ is a diagonal or anti-diagonal contour such
as the ones depicted in Fig.~\ref{fig:1}.
In the $\sigma_i^y$'s eigenbasis, the parities of the number of spin flips
along such contours are thus conserved.
This important property allows for ED  of "rather
large" systems  up to $N=32$ spins, with the periodic boundary conditions defined in Ref.~\onlinecite{Kitaev03}. The product of two parity operators defined on parallel contours is furthermore equal to
the product of all $A_s$ and $B_p$ operators between the corresponding contours, which relates parities of magnons to that of anyons.

%
%
\emph{Self-duality ---}
%
%
This correspondence is only one signature of the strong link between both types of QP which roots in
a crucial property of the model~: its self-duality. This  feature directly stems
from the mapping of the transverse-field TCM onto the Xu-Moore model which is 
self-dual~\cite{Xu04,Xu05,Chen07_1}. Indeed, let us introduce spin variables living on the dual lattice $\widetilde \Lambda$ (see Fig.~\ref{fig:1})
%
%
\begin{equation}
  \label{eq:mapping_xm}
  \widetilde \sigma_{j_s}^z=A_s, \quad \widetilde \sigma_{j_p}^z=B_p, \quad \mbox{and} \quad
  \widetilde \sigma_j^x=\prod_{j>i} \sigma_i^y,
\end{equation}
%
%
where $j_{s(p)}$ denotes the center of a star (plaquette).
The notation $j>i$ defines the set of all sites $i \in \Lambda$ whose two coordinates are smaller than those of $j\in \widetilde \Lambda$. \mbox{Hamiltonian (\ref{eq:ham})} can then be rewritten as that of the Xu-Moore model
%
%
\begin{equation}
  H = - J \sum_{j} \widetilde \sigma_j^z
  - h_y \sum_{\widetilde p} \prod_{j \in \widetilde p} \widetilde\sigma_j^x,
\end{equation} 
%
%
where the first (second) sum is performed over all sites $j$
(plaquettes $\widetilde p$) of $\widetilde \Lambda$.
Note that the above mapping only holds in the thermodynamic limit and for open
boundary conditions, and that the infinite number of spins involved in the
definition of $\widetilde \sigma_j^x$ cannot keep track of degeneracies.
In particular, ED spectra of the transverse-field TCM with periodic boundary conditions discussed below are not symmetric under the exchange $h_y \leftrightarrow J$.

%
%
\begin{figure}[t]
 \includegraphics[width=0.83\columnwidth]{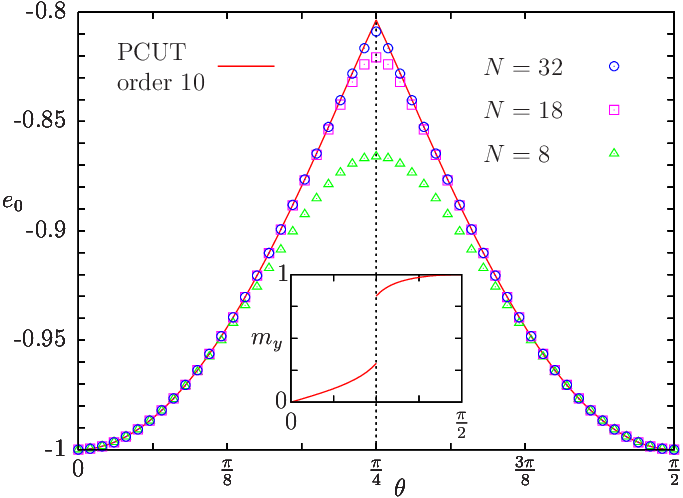}
 \caption{(Color online). Ground-state energy per spin $e_0$ obtained from PCUT and ED. 
   The vertical dashed line marks the transition at $\theta=\pi/4$.
   {\it Inset}:~magnetization $m_y=-\partial_{h_y} e_0$.}
 \label{fig:2}
\vspace{-10pt}
\end{figure}
%
%

Interestingly, Nussinov and Fradkin~\cite{Nussinov05_1} showed that the Xu-Moore
model can also be mapped onto the quantum compass model~\cite{Kugel82}. This model has focused much attention recently and latest numerical results plead in favor of a unique first order transition at the self-dual point~\cite{Dorier05,Chen07_3,Orus09_1} contrary to the original claim by Xu and Moore~\cite{Xu04,Xu05}. 
This scenario that we shall confirm in the following immediately implies that the topological phase is rather well protected from transverse fields compared to parallel fields. Indeed, in the former case it breaks down at $h_y=J$ whereas in the latter case, the transition takes place for a field magnitude of order $J/3$~\cite{Vidal09_1}.

%
%
\emph{Perturbative analysis ---}
%
%
As in Ref.~\onlinecite{Vidal09_1}, a PCUT treatment can be set up in the
limit of low (high) field, highlighting the role of the corresponding
QP, namely anyons (magnons). Basically, this method transforms $H$ into an effective Hamiltonian unitarily equivalent to $H$ but conserving the number of QP~\cite{Wegner94,Stein97,Knetter00_1,Knetter03_1,Vidal08_2,Vidal09_1}.
It allows to investigate the thermodynamic limit and sorts the energy levels according to their number of QP. 
We shall thus investigate low-energy sectors and confront the QP interpretation stemming from PCUT with ED spectra.

%
%
 \emph{0QP sector ---}
%
%
By construction, the 0QP state is  the ground-state and lies in the
symmetry subspace where all parities are even. We have computed the perturbative
expansion of the ground-state energy per spin $e_0$, up to order 10, in the low-field regime. Setting $J=\cos\theta$, \mbox{$h_y=\sin\theta$}, and $t=\tan\theta$, it reads as
%
%
\begin{eqnarray}
  \label{e0}
  \frac{e_0}{\cos \theta}&=&-1-\frac{t^2}{8}-\frac{13\, t^4}{1536}
  -\frac{197\, t^6}{98304}-\frac{163885\, t^8}{226492416} \nonumber \\
&&  -\frac{186734746441\, t^{10}}{587068342272000}.
\end{eqnarray}
%
%
This formula and its high-field counterpart (obtained by exchanging $h_y$ and
$J$) are represented in Fig.~\ref{fig:2}. As can be seen, the agreement between (\ref{e0}) and ED results for $N=32$ spins is remarkable. Let us mention that the PCUT series expansion is rather well converged, since the difference between order 8 and 10 is of the order $10^{-4}$ for all values of $\theta$.
Furthermore, a Pad\'e approximant analysis gives $e_0(\theta=\pi/4)=-0.8038(1)$,
which perfectly matches previous numerical results~\cite{Dorier05,Orus09_1}.

The cusp in the ground-state energy at $\theta=\pi/4$ indicates that the topological phase breaks
down when $h_y$ reaches the value $J$, in agreement with the self-duality of
the model. The transition point is best detected when looking at the
magnetization in $y$-direction, which is obtained from the Hellmann-Feynman
theorem~: $m_y=-\partial_{h_y} e_0$. It displays a jump that reveals the
first order nature of the transition (see inset of Fig.~\ref{fig:2}).  Our
perturbative treatment therefore confirms the order of the transition in the quantum compass model~\cite{Dorier05, Chen07_3,Orus09_1}.

\begin{figure}[t]
  \includegraphics[width=0.83\columnwidth]{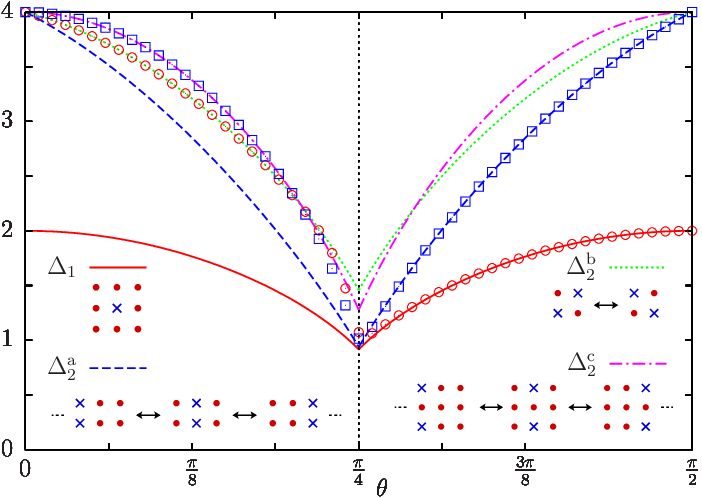}
  \caption{(Color online). Comparison between PCUT and ED results ($N=32$) for the 1QP gap $\Delta_1$ and lowest 2QP gaps $\Delta_2^{\mathrm{a,b,c}}$. Pictograms give an illustration of the four corresponding  states (with $\pi/4$-tilted lattice compared to Fig.~\ref{fig:1}). Crosses denote particles and filled circles empty sites.}
 \label{fig:3}
\end{figure}

%
%
\emph{1QP  sector ---}
%
%
We now turn to the properties of a single QP. These excitations are static (dispersionless) due to parity conservation.
In the high-field phase, they belong to sectors with all spin-flip parities even,
except exactly one diagonal and one anti-diagonal parities, crossing at the
QP's position. We computed the energy gap $\Delta_1$ of this 1QP sector which, at order 10 and in the low-field regime, reads as
%
%
\begin{eqnarray}
  \label{d1}
  \frac{\Delta_1}{\cos\theta}&=&2-\frac{t^2}{2}-\frac{15\, t^4}{128}
  -\frac{575\, t^6}{12288}-\frac{26492351\, t^8}{1019215872} \nonumber \\
  &&-\frac{185172052871 \, t^{10}}{24461180928000}.
\end{eqnarray}
%
%
Once again, ED data perfectly match analytical results, as can be
inferred from Fig.~\ref{fig:3}, whose upper left pictogram gives a
representation of a 1QP state.  At the transition point $h_y=J$, Pad\'e extrapolations lead to
$\Delta_1(\theta=\pi/4)=0.9005(1)$. 
As already mentioned, eigenstates on both sides of the transition have to be interpreted
either in terms of dressed anyons or in terms of dressed magnons. Furthermore, we remark that ED, performed on clusters with periodic boundary conditions, can only detect the 1QP excitation at large field since the excitation of a single anyon in the topological phase is forbidden for such boundary conditions. 
We also note that the one-magnon state is connected to one of the two-anyon states which we study below.
%
%
%
%

%
%
\emph{2QP sector ---}
%
%
In the high-field phase, parity symmetries of a two-magnon state can be of two kinds. 
The first is obtained by setting all parities even except two diagonal {\it or}
anti-diagonal parities which are odd as in the two lowest pictograms in Fig.~\ref{fig:3}. In this case, the two magnons can only move in the direction orthogonal to their relative position. This is a nice illustration of the  dimensional reduction phenomenon~\cite{Xu04,Xu05} in which the transverse magnetic field induces a one-dimensional correlated hopping and leads to the formation of bound states.
The second kind is obtained by setting all parities even except
two diagonal {\it and} two anti-diagonal parities which are odd, as depicted for instance
in the upper right pictogram in Fig.~\ref{fig:3}. In such a configuration, 
parity conservation imposes very limited kinetics of the two magnons.  Again, the transverse magnetic field leads to strong binding effects.

The self-duality allows for a similar analysis in the topological phase for anyons, with the restriction that only two-charge or two-flux states are allowed for a system with periodic boundary conditions. 
Such excitations have bosonic statistics. However, dyonic bound states made of one charge and one flux (with fermionic statistics) only exist for open boundary conditions.
We have calculated all 2QP excitation energies up to order 8. Hereafter, we 
provide the three lowest gaps in the low-field phase corresponding to the three 2QP configurations shown in the pictograms of Fig.~\ref{fig:3},
%
%
\begin{eqnarray}
  \frac{\Delta_2^{\mathrm{a}}}{\cos\theta} &=&
  4-2 t-\frac{t^2}{2}+\frac{t^3}{16}-\frac{17\, t^4}{96}
  +\frac{337\, t^5}{14144}-\frac{1895\, t^6}{18432}\nonumber \\
  &&+\frac{236471\, t^7}{4718592}-\frac{386712919\, t^8}{5096079360},
  \label{da2}\\
  \frac{\Delta_2^{\mathrm{b}}}{\cos\theta} &=&
  4-t-\frac{5\, t^2}{8}+\frac{t^3}{32}-\frac{353\, t^4}{1536}
  +\frac{1355\, t^5}{36864} \\
  &&-\frac{247511\, t^6}{1769472}+\frac{43261\, t^7}{1048576}
  -\frac{1906002767\, t^8}{20384317440},\label{db2} \nonumber \\
  \frac{\Delta_2^{\mathrm{c}}}{\cos\theta} &=&
  4-2t^2-\frac{t^4}{24}-\frac{1845\, t^6}{16384}
  -\frac{200004589\, t^8}{5096079360}. \label{dc2}
\end{eqnarray}
%
%

These PCUT series,  plotted in Fig.~\ref{fig:3}, seem to indicate that $\Delta_1$
and $\Delta_2^{\mathrm{a}}$ are equal at the transition.
As previously, we compared them with ED data which, for $N=32$ spins, imply to deal with blocks containing up to 16 million states.
Although ED and PCUT results almost lie on top of each other, ED spectra reveal the formation of
energy jumps at the transition point, which can only be explained by level
crossings with higher-energy states occurring in the thermodynamical limit.
These crossings cannot be captured by the PCUT approach, whose perturbative nature
imposes an adiabatic continuation of levels. However, we insist on the validity of our results for all $\theta$'s, except at the transition point.

We therefore conclude that the level crossing responsible for the cusp in the ground-state energy does not originate from 1QP and 2QP levels, since these excitation energies are finite at  $\theta=\pi/4$.

\begin{figure}[t]
  \includegraphics[width=0.83\columnwidth]{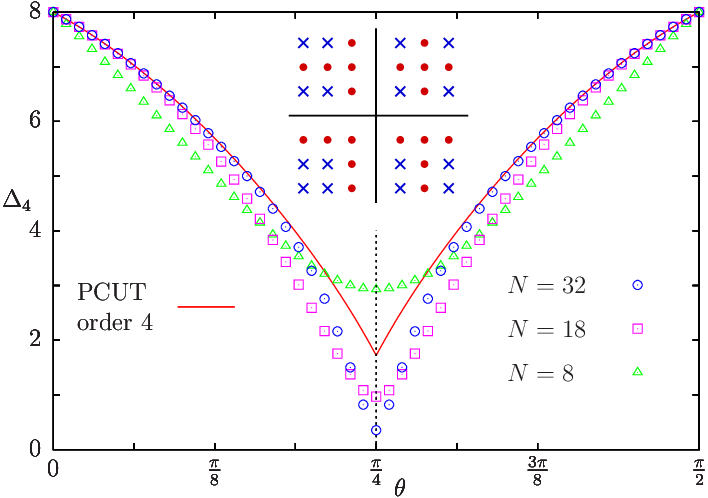}
  \caption{(Color online). Lowest 4QP gap $\Delta_4$ obtained from PCUT and  ED. Pictograms
  illustrate four 4QP configurations, with different relative positions.}
  \label{fig:4}
\end{figure}

%
%
 \emph{4QP sector ---}
%
%
To address the origin of the cusp, we now look for the lowest excited state belonging to the
same symmetry sector as the ground state. PCUT energy ordering suggests a 4QP state as a natural candidate.
Such a 4-magnon state (or a two-flux and two-charge state with bosonic statistics) is built from all configurations where the magnons occupy the corners of a rectangle.
These can be linked to the configuration where the four QP form a close-packed square, by shifting the center of mass and/or the relative positions of the QP. Four such configurations are shown in Fig.~\ref{fig:4}.

%
%
%
%

In contrast to 2QP states, 4QP states in this parity sector have a two-dimensional dispersion. However, 
a partial dimensional reduction still occurs for the relative motion of the QP. Indeed, the corresponding
effective Hamiltonian at order $n$ in perturbation is found to be that of a single particle moving in $n$ coupled one-dimensional chains, with an impurity whose extension grows with $n$ (details will be given elsewhere). 
As it frequently occurs in this type of problem, the bound state associated to this imurity cannot be obtained perturbatively, and one has to resort to a numerical diagonalization of the effective Hamiltonian.

The gap $\Delta_4$ of the 4QP bound state, obtained from the fourth-order effective Hamiltonian, is shown in Fig.~\ref{fig:4}, together with ED results. 
Both match away from the transition but, contrary to the ED gap which goes to zero at the transition, the PCUT gap remains finite. At order 3, one gets $\Delta_4= 1.728$ at $\theta=\pi/4$ whereas at order 4,  $\Delta_4=1.721$ suggesting a fast convergence. 
This finite value shows once again that PCUT miss level crossings occurring at the self-dual point, but give reliable results everywhere else. This discrepancy can be readily explained by the first order transition, in which a cascade of level crossings from high-energy states down to the ground state occurs in the thermodynamic limit.

%
%
\emph{Perspectives ---}
%
%
We have shown that a transverse magnetic field in the TCM is the source of important
binding effects, leading to a sequence of bound states with reduced dimensional kinetics, in deep contrast with the parallel field case~\cite{Vidal09_1}. The fate of these bound states in a general magnetic field, where single-quasiparticle excitations are also dispersive, is a fascinating issue left for future studies.

\acknowledgments We thank B. Dou\c{c}ot for discussions and J. Dorier for sharing his numerical data~\cite{Dorier05}. K.P.S. acknowledges ESF and EuroHorcs for funding through his EURYI.


%

\end{document}